\documentclass[pss]{wiley2sp} % provides pss two-column style
\usepackage{amsmath}
\usepackage{bm}              % uncomment these two packages if you
\usepackage{w-greek}         % need extended greek-letter functionality in math mode

\newcommand{\dbraket}[1]{\ensuremath{ \langle \! \langle #1 \rangle \! \rangle}}

\begin{document}

% Title of the article
\title{Violation of detailed balance for charge-transfer statistics in Coulomb-blockade systems}

% Abbreviated title for the page headers
\titlerunning{Violation of detailed balance for charge-transfer statistics in Coulomb-blockade systems}

% Authors
\author{%
  Philipp Stegmann\textsuperscript{\Ast,\textsf{\bfseries 1}} and
  J\"urgen K\"onig\textsuperscript{\textsf{\bfseries 1}}}

% Abbreviated list of authors for the page headers
\authorrunning{P. Stegmann and J. K\"onig}

%E-mail-address of corresponding author
\mail{e-mail
  \textsf{philipp.stegmann@uni-due.de}, Phone:
  +49-203-379-3323, Fax: +49-203-379-3665}

% author's affiliations/addresses
\institute{%
  \textsuperscript{1}\,Theoretische Physik, Universit\"at Duisburg-Essen and CENIDE, 47048 Duisburg, Germany}
  
%\received{XXXX, revised XXXX, accepted XXXX} % do not change, will be filled in by the publisher
%\published{XXXX} % do not change, will be filled in by the publisher

% Please select about four verbal keywords for your manuscript.

\abstract{%
% This is a macro for the typesetting of two-column text in an
% abstract. It will typeset the two arguments in \abstcol{}{} as the
% left and right column inside the abstract box. At the
% columnbreak there will be always a columnbreak (\par), so both
% columns start with a new paragraph. No automatic column height
% balancing is done.
%
% If used with a \titlefigure it will silently output both
% parameters as consecutive paragraphs.
%
% The macro is defined exclusively inside the argument of \abstract{};
% if used outside it will raise an error.
%
% Usage: \abstcol{<left column>}{<right column>}
\abstcol{%
We discuss the possibility to generate in Coulomb-blockade systems steady states that violate detailed balance.
This includes both voltage biased and non-biased scenarios.
The violation of detailed balance yields that the charge-transfer statistics for electrons tunneling into}
{an island experiencing strong Coulomb interaction is different from the statistics for tunneling out.
This can be experimentally tested by time-resolved measurement of the island's charge state.
We demonstrate this claim for two model systems.
}}
% The class file requires the standard graphicx Latex package. See the 'LaTeX
% standard graphics and color packages documentation' for more information at
% <http://tug.ctan.org/tex-archive/macros/latex/required/graphics/grfguide.pdf>.
%
% Accepted figure file formats depend on which LaTeX flavour is used.
% Classic LaTeX is always able to use Encapsulated Postscript (EPS);
% PDFLaTeX can't use this but accepts PDF, JPG, PNG, and GIF formats.
%
% See examples for implementing graphics in floating figure environments later in this file.
% If \titlefigure is given, it takes as its mandatory parameter the
% name (without extension) of some figure file.

\maketitle   % please do not remove

\section{Introduction}
Tunneling of electrons into and out of Coulomb-blockade devices such as semiconductor quantum dots or metallic single-electron boxes is a stochastic process.
The dynamics of the system is governed by the tunneling rates for the transition between different states of the confined region (quantum dot or single-electron box), in the following referred to as the island.
The island state is characterized by its total charge but may also possess other degrees of freedom such as spin.
The tunneling rates enter a rate equation that can be used to calculate the time average of quantities such as the island charge, the current through the island, or the current noise.

However, recent progress in nanotechnology has made it possible to monitor tunneling into and out of individual electrons in a time resolved manner.
For this, the island has been electrostatically coupled to a sensitive electrometer.
In the case of quantum dots, the current through an electrostatically-coupled quantum point contact indicates the island's charge state~\cite{gustavsson_counting_2005,fujisawa_bidirectional_2006,gustavsson_measurements_2007,fricke_bimodal_2007,flindt_universal_2009,gustavsson_electron_2009,fricke_high_2010,fricke_high-order_2010,komijani_counting_2013}. 
Similarly, the charge state of a metallic single-electron box can be detected via the current through an electrostatically-coupled single electron 
transistor~\cite{martinis_metrological_1994,dresselhaus_measurement_1994,lotkhov_storage_1999,lu_real_2003,bylander_current_2005}.
Also other, e.g., optical~\cite{kurzmann_optical_2016} or interferometric~\cite{dasenbrook_dynamical_2016}, detection schemes are conceivable. 
In all cases, the measured time trace of the detector allows for studying the full counting statistics of the charge transfer between island and leads. Counting statistics has recently been utilized to feedback-control quantum transport~\cite{poeltl_feedback_2011,daryanoosh_stochastic_2016,wagner_squeezing_2016} or to study non-Markovian~\cite{braggio_full_2006,flindt_counting_2008,flindt_counting_2010} and frequency-dependent~\cite{emary_frequency-dependent_2007,ubbelohde_measurement_2012} processes. Moreover, the impact of ferromagnetic leads~\cite{lindebaum_spin-induced_2009}, electron-phonon interactions~\cite{schmidt_charge_2009,souto_transient_2015}, and the influence of Andreev reflections~\cite{cuevas_full_2003,johansson_full_2003,pilgram_noise_2005,morten_full_2008} have been investigated.
\begin{figure*}[t]
\begin{center}
	\includegraphics[scale=1.0]{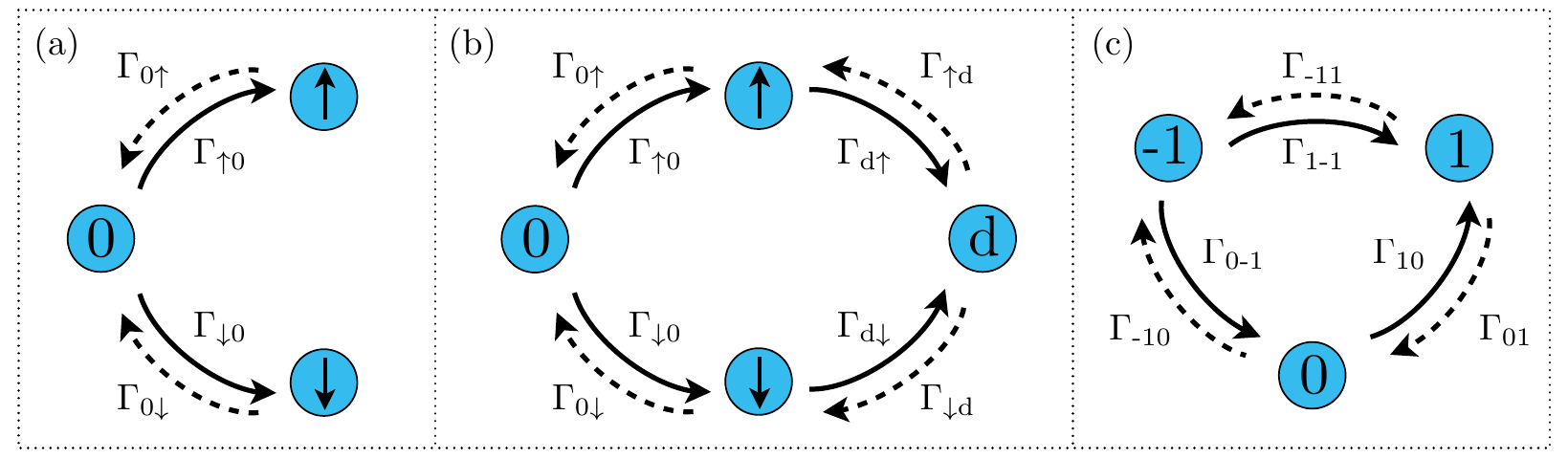}
	\caption{
Three examples of a stochastic model realized in Coulomb-blockade systems.
Solid (dashed) lines indicate transitions that are counted for the statistics of tunneling in (out).
A violation of detailed balance, indicated by different counting statics for tunneling in and tunneling out, is possible for model (b) and (c) but not for (a).}	  
\label{figure1}
\end{center}
\end{figure*}

Throughout this paper, we only consider steady-state situations, i.e., transient behavior after switching on the device is excluded and counting in the detector starts only after device plus detector have reached their steady state.
But we allow for both voltage biased and non-biased scenarios.
In the latter case, only one lead is coupled to the island, in the former case, a bias voltage is applied to two leads that drive a current through the island. 
It is the goal of this paper to show that, in both cases it is possible to generate steady states that violate detailed balance. 
We illustrate for two specific model systems how the violation of detailed balance is detectable in the tunneling statistics.
The issue of detailed-balance violation has recently attracted interest in the statistical-mechanics community \cite{Zia_possible_2006,Zia_probability_2007}.
Stationary states that violate detailed balance are not fully characterized by the probability distribution alone since closed-loop probability currents occur.
This has important implications for computer simulations of stochastic systems.
Furthermore, the connection between closed-loop probability currents and entropy production has been investigated.

\section{Stochastic Systems}
A stochastic system is described by the rate equation
\begin{equation}
\label{master}
	\dot p_\chi = \sum_{\chi'\neq \chi} \left( \Gamma_{\chi \chi'} p_{\chi'} - \Gamma_{\chi' \chi} p_{\chi} \right)
\, ,
\end{equation}
where $p_{\chi}$ is the probability of the system to be in state $\chi$ and $\Gamma_{\chi'\chi}$ is the transition rate from state $\chi$ to $\chi'$.
In the steady-state limit, the probability distribution $\{ p_\chi \}$ is time independent, i.e., the time derivative $\dot p_\chi$ vanishes and the sum on the right-hand side of Eq.~(\ref{master}) is zero.

\subsection{Detailed Balance}
The stationary state is said to obey detailed balance if not only the {\it sum} but {\it each summand} on the right-hand side of Eq.~(\ref{master}) vanishes,
\begin{equation}\label{detailedB}
	\Gamma_{\chi \chi'} p_{\chi'} = \Gamma_{\chi' \chi} p_{\chi} \, ,	
\end{equation}
for each pair $\chi, \chi'$~\cite{Zia_possible_2006,Zia_probability_2007}.
The combination $\Gamma_{\chi' \chi} p_{\chi}$ can be interpreted as the probability current for the system to flow from state $\chi$ to $\chi'$. 
Detailed balance, then, means that the probability current from $\chi$ to $\chi'$ minus the one from $\chi'$ to $\chi$ is zero, i.e., the net probability current is zero $I_{\chi'\chi}=\Gamma_{\chi' \chi} p_{\chi}-\Gamma_{\chi \chi'} p_{\chi'}=0$
\cite{note1}.

To illustrate its implications, let us consider the following three stochastic systems that will be relevant for the discussion in the second part of the paper.
The first one, sketched in Fig.~\ref{figure1}a, consists of three states labeled by $\chi=0, \uparrow, \downarrow$, and transitions are allowed between $0$ and $\uparrow$ as well as between $0$ and $\downarrow$ but not directly between $\uparrow$ and $\downarrow$.
In this case, the stationary state must obey detailed balance, as otherwise the net probability current $I_{\uparrow 0}$ or $I_{\downarrow 0}$ would be non-zero, which is incompatible with the steady-state condition.

Next, we extend the model by adding a fourth state d with transition rates from and to the states $\uparrow$ and $\downarrow$, see Fig.~\ref{figure1}b.
Depending on the values of the rates, this system can accommodate steady states that do not obey detailed balance.
In such a state, there is a finite net probability current flowing around the loop $0 \rightarrow \, \uparrow \, \rightarrow {\rm d} \rightarrow \, \downarrow \, \rightarrow 0$ (or in the opposite direction $0 \rightarrow \, \downarrow \, \rightarrow {\rm d} \rightarrow \, \uparrow \, \rightarrow 0$).

Detailed-balance violating steady states are already possible for stochastic systems consisting of three states, but only if direct transitions between all the states $-1$, $0$, $1$ are allowed, see Fig.~\ref{figure1}c, and the tunneling rates are chosen properly. 

\subsection{Full Counting Statistics}
In Coulomb-blockade devices, steady states that violate detailed balance can be detected by analyzing the charge-transfer statistics.
The latter is accessible by a time-resolved measurement of the island's charge state. 
Let us assume that, in the models sketched in Figs.~\ref{figure1}a and \ref{figure1}b, the detector can distinguish between a quantum dot being empty, $0$, being singly occupied with $\uparrow$ or $\downarrow$ (without being sensitive to the spin degree of freedom), and being doubly occupied, d.
In the model sketched in Fig.~\ref{figure1}c, we assume the detector to distinguish between three different charge states $-1$, $0$, and $1$ (relative to some reference charge).

In the time trace of the detector current one can identify all tunneling events in which an electron has tunneled {\it into} the island (solid lines in Fig.~\ref{figure1}).
By slicing the full time trace into intervals of length $t$ each, one can obtain the distribution of the probabilities $P^{\rm in}_N(t)$ that $N$ (with $N\ge 0$) electrons have tunneled into the island within a time span of length $t$. 
Alternatively, one may decide to count only electrons tunneling {\it out} (dashed lines in Fig.~\ref{figure1}) to obtain the distribution $P^{\rm out}_N(t)$.
Detailed balance implies that the two distributions for tunneling in and tunneling out are identical, since then the probability to find a given sequence of island states, $\chi_1 \rightarrow \chi_2 \rightarrow \ldots \rightarrow \chi_k$, is the same as finding the opposite sequence, $\chi_k \rightarrow \ldots \rightarrow \chi_2 \rightarrow \chi_1$.
Therefore, a difference between $P^{\rm in}_N(t)$ and $P^{\rm out}_N(t)$ signals the violation of detailed balance.
However, the reverse statement is not necessarily true: a violation of detailed balance may or may not yield a difference between $P^{\rm in}_N(t)$ and $P^{\rm out}_N(t)$, depending on which transitions are counted. 

A convenient way to quantify any full counting statistics $P_N(t)$ is to use moments or cumulants.
Since, in our case, the stochastic variable $N$ is an integer rather than a continuous variable, it is advantageous to employ so-called {\it factorial} moments, i.e., moments $\langle N^{(m)}\rangle (t) := \sum_N N^{(m)} P_N(t)$ of the factorial power $N^{(m)}:= N(N-1)\ldots (N-m+1)$ rather than the ordinary power $N^m$ used for ordinary moments.
Factorial moments are derived from the generating function
\begin{equation}
	{\cal M}_{\rm F}(z,t) := \sum_N (z+1)^N P_N(t) \, ,	
\end{equation}
via the derivatives ${\langle {N^{(m)}}\rangle} (t)= \partial_z^m {\cal M}_{\rm F}(z,t)|_{z=0}$ taken at $z=0$.
The corresponding factorial cumulants $C_{{\rm F},m}(t):= {\dbraket {N^{(m)}}} (t)$ are obtained from 
\begin{equation}
\label{cumulants}
	C_{{\rm F},m}(t) := \frac{\partial^m}{\partial z^m} \ln {\cal M}_{\rm F}(z,t) \Bigl |_{z=0} \, .
\end{equation}
In the context of single-electron tunneling, the use of factorial rather than ordinary cumulants has been suggested~\cite{kambly_factorial_2011,kambly_time-dependent_2013} as a convenient tool to identify interactions in the system. 
Furthermore, factorial and generalized factorial cumulants~\cite{stegmann_2015} have been analyzed in the short-time limit to detect the presence of fundamental tunneling processes (such as Andreev tunneling) of two electrons simultaneously~\cite{stegmann_2016}. 
Here, we are going to use factorial cumulants to probe the violation of detailed balance by comparing $C_{{\rm F},m}^{\rm in}(t)$ with $C_{{\rm F},m}^{\rm out}(t)$.

To evaluate the generating function for a given stochastic system, we follow along the lines described in Ref.~\cite{stegmann_2015}. First, we construct a matrix ${\bf W}^{\rm in}_{z}$ in which the matrix element $\left({\bf W}^{\rm in}_{z} \right)_{\chi'\chi}$ is associated with the transition from state $\chi$ to state $\chi'$. The matrix element is given by the transition rate $\Gamma_{\chi'\chi}$ times $z^k$ where $k$ is number of electrons having entered the island due to the transition. For example, the stochastic model depicted in Fig.~\ref{figure1}b yields
\begin{equation}\label{Winmat}
\begin{split}
	&{\bf W}^{\rm in}_{z} = \\
	&\left( \begin{array}{cccc}
		-\Gamma_{\uparrow 0}-\Gamma_{\downarrow 0} & \Gamma_{0 \uparrow} & \Gamma_{0 \downarrow} & 0 
\\
		z \Gamma_{\uparrow 0} & -\Gamma_{0 \uparrow}-\Gamma_{{\rm d}\uparrow}  & 0& \Gamma_{\uparrow {\rm d}} 
\\
		z \Gamma_{\downarrow 0} & 0 & -\Gamma_{0 \downarrow}-\Gamma_{{\rm d} \downarrow} & \Gamma_{\downarrow {\rm d}} 
\\
		0 & z \Gamma_{{\rm d} \uparrow} & z \Gamma_{{\rm d} \downarrow} & -\Gamma_{\uparrow {\rm d}}-\Gamma_{\downarrow {\rm d}} 
	\end{array} \right)\, ,
\end{split}
\end{equation}
where rows and columns are arranged to correspond to the states $0,\uparrow, \downarrow,{\rm d}$. In the given example, no term with $z^2$ appears since [in contrast to the stochastic model shown in Fig.~\ref{figure1}c] there is no transition for which the transfer of two electrons is counted. As shown in Ref.~\cite{stegmann_2015}, the generating function yielding $C^{\rm in}_{{\rm F}, m}$ via Eq.~(\ref{cumulants}) can, then, be expressed as
\begin{equation}
{\cal M}^{\rm in}_{\rm F}(z,t)={\bf e}^T \cdot \exp({\bf W}^{\rm in}_{z+1}t){\bf P}_{\rm stat}\, ,
\end{equation}
with the vector ${\bf e}^T = (1,1, \ldots , 1)$. The stationary probability distribution is given by ${\bf W}^{\rm in}_1 {\bf P}_{\rm stat}=0$ and ${\bf e}^T\cdot {\bf P}_{\rm stat} =1$.

For the generating function ${\cal M}^{\rm out}_{\rm F}(z,t)$ of $C^{\rm out}_{{\rm F}, m}$, the matrix ${\bf W}^{\rm in}_z$ must be replaced by ${\bf W}^{\rm out}_z$. Its matrix elements are constructed in a identical way as described above, except that $k$ is the number of electrons having left the island due to the transition. For example, the matrix ${\bf W}^{\rm out}_z$ of the stochastic model depicted in Fig.~\ref{figure1}b is obtained form Eq.~(\ref{Winmat}) if the factors $z$ are removed from the lower-left off-diagonal matrix elements and put, instead, to the upper-right terms.

\begin{figure}[t]
\begin{center}
\includegraphics[scale=0.9]{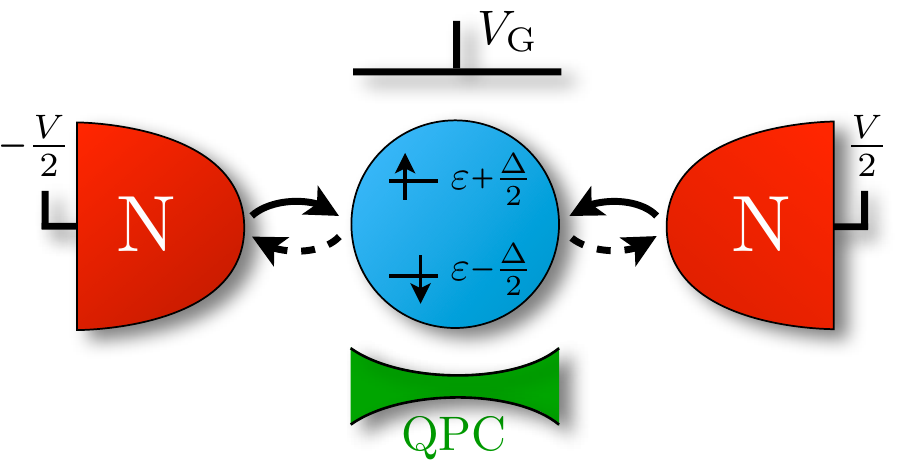}
	\caption{
A single-level quantum dot (blue) subjected to a Zeeman field is tunnel coupled to two metallic leads (red). A bias voltage~$V$ is applied between the leads and the gate voltage~$V_\text{G}$ tunes the level position~$\varepsilon$. The current through a quantum point contact (green) monitors the dot charge as function of time.
	}	  
\label{figure2}
\end{center}
\end{figure}

\section{Single-level quantum dot with Zeeman field}
\begin{figure*}[t]
{\includegraphics[scale=1.00]{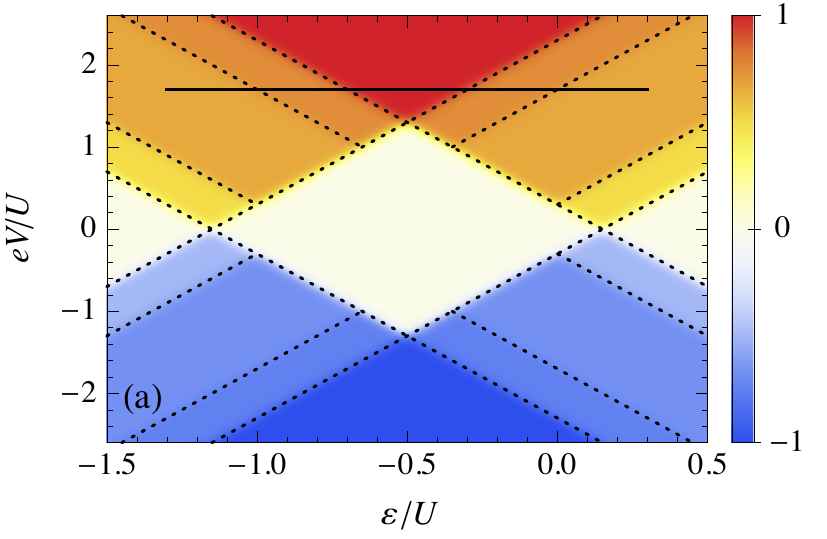}}
{\includegraphics[scale=1.00]{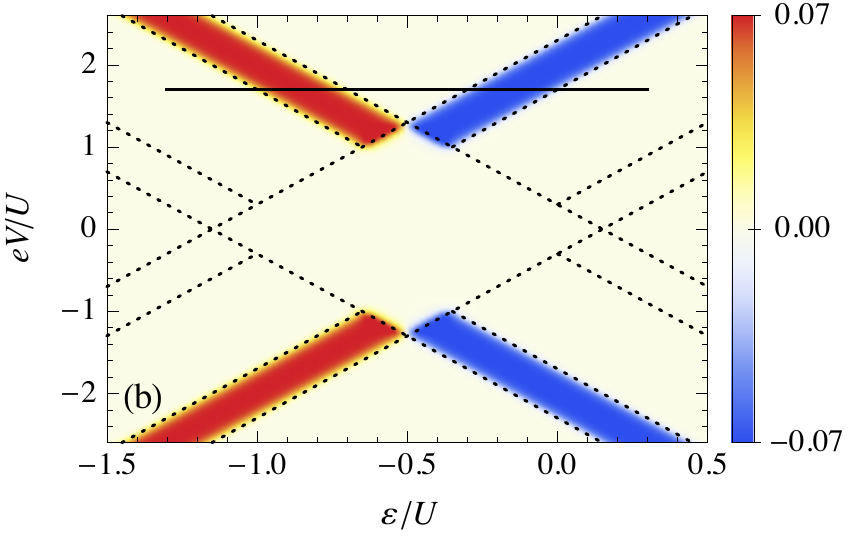}}
\caption{
(a) Current in units of $e\Gamma_{\rm L}$ through the quantum-dot system depicted in Fig.~\ref{figure2}.
(b) The difference $C_{{\rm F},2}^{\rm in}-C_{{\rm F},2}^{\rm out}$ of the second factorial cumulant as function of the level energy~$\varepsilon$ and bias voltage~$V$. The parameters are $\Delta=0.3 \,U$, $k_{\rm B}T=0.02 \,U$, $\Gamma_{\rm L}=\Gamma_{\rm R}$, and $t=2/\Gamma_{\rm L}$. The dotted lines mark positions of the resonances of the quantum-dot excitation energies with the the Fermi level of the source or drain electrode. A nonvanishing value of $C_{{\rm F},2}^{\rm in}-C_{{\rm F},2}^{\rm out}$ in (b) indicates the violation of detailed balance.
	}
\label{figure3}
\end{figure*}
As a first example, we consider a single-level quantum dot subject to a magnetic field and weakly tunnel coupled to two leads $r=\text{L},\text{R}$ with tunnel-coupling strength $\Gamma_r$, see Fig.~\ref{figure2}.
The Zeeman field $\Delta$ splits the orbital level $\varepsilon$ into $\varepsilon_\uparrow= \varepsilon + \Delta/2$ and $\varepsilon_\downarrow= \varepsilon - \Delta/2$ for spin $\sigma=\uparrow,\downarrow$.
The charging energy for double occupancy is denoted by $U$.
A bias voltage $V$ is symmetrically applied to the two leads, described by the electrochemical potentials $\mu_{\rm L} = eV/2$ and $\mu_{\rm R} = -eV/2$ for the left and right lead, respectively.
Fermi's golden rule yields tunneling rates
\begin{eqnarray}
	\Gamma_{\sigma 0} &=& \sum_{r={\rm L,R}} \Gamma_r f_r(\varepsilon_\sigma)\, ,
\\
	\Gamma_{0 \sigma} &=& \sum_{r={\rm L,R}} \Gamma_r [1-f_r(\varepsilon_\sigma)]\, ,
\\
	\Gamma_{{\rm d} \sigma} &=& \sum_{r={\rm L,R}} \Gamma_r f_r(\varepsilon_{\bar \sigma} + U)\, ,
\\
	\Gamma_{\sigma {\rm d}} &=& \sum_{r={\rm L,R}} \Gamma_r [1-f_r(\varepsilon_{\bar \sigma} + U)] 
\, .
\end{eqnarray}
Here, we made use of the Fermi function $f_r(x) =1/ \{\exp[(x-\mu_r)/k_{\rm B}T]+1\}$ and $\bar\sigma$ is the spin opposite to $\sigma$. 
 \begin{figure}[b]
	\includegraphics[scale=1.00]{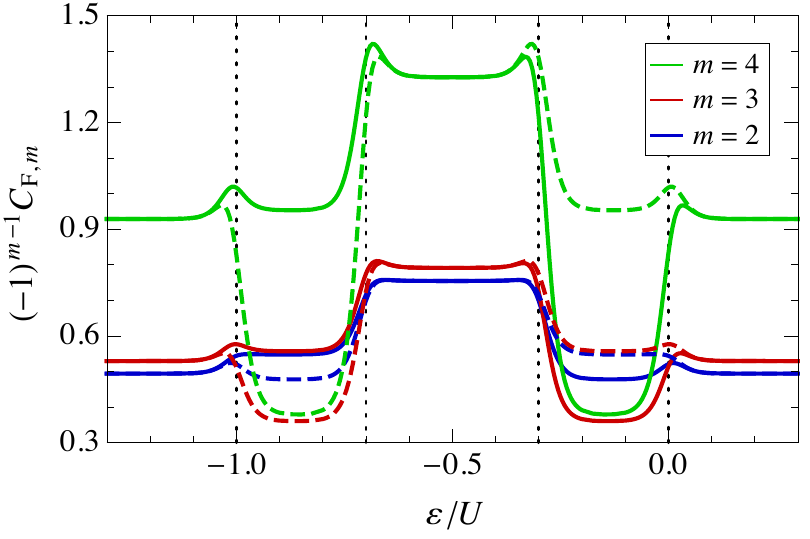}
	\caption{
Factorial cumulants of tunneling in $C_{{\rm F},m}^{\rm in}$ (solid lines) and tunneling out events $C_{{\rm F},m}^{\rm out}$ (dashed lines). The parameters are the same as in Fig.~\ref{figure3} with $eV=1.7\,U$.
The dotted vertical lines indicate the positions of the resonances of the quantum-dot excitation energies with the the Fermi level of the source or drain electrode.
	}	  
\label{figure4}
\end{figure}
\begin{figure}[b]
\begin{center}
	\includegraphics[scale=1.36]{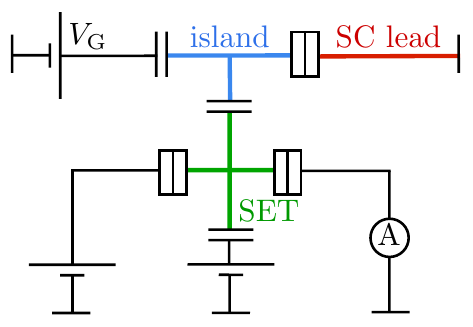}
	\caption{
A normal-state metallic island (blue) tunnel coupled to one superconducting lead (red). Via the gate voltage $V_{\rm G}$, the gate charge $n_\text{G}$ of the island can be tuned. The current through a single-electron transistor (green) monitors the island charge state as function of time.
	}	  
\label{figure5}
\end{center}
\end{figure}
\begin{figure*}[t]
{\includegraphics[scale=1.00]{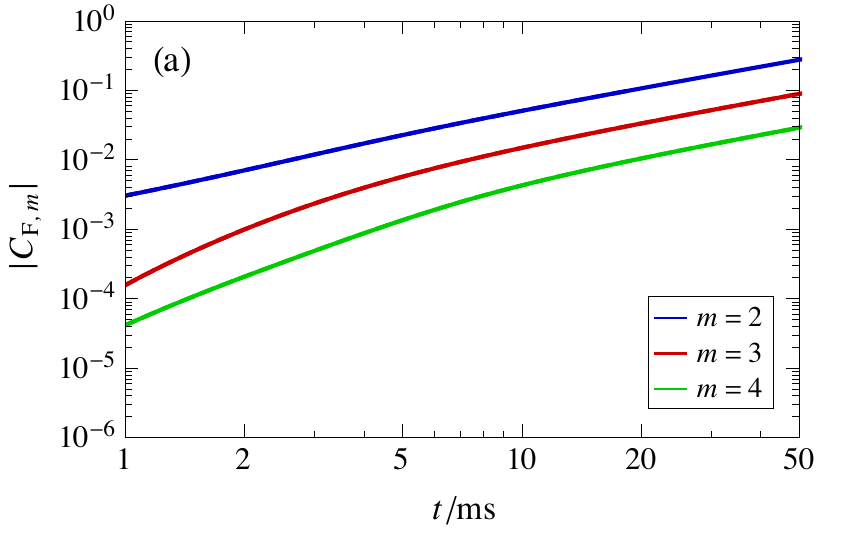}}
{\includegraphics[scale=1.00]{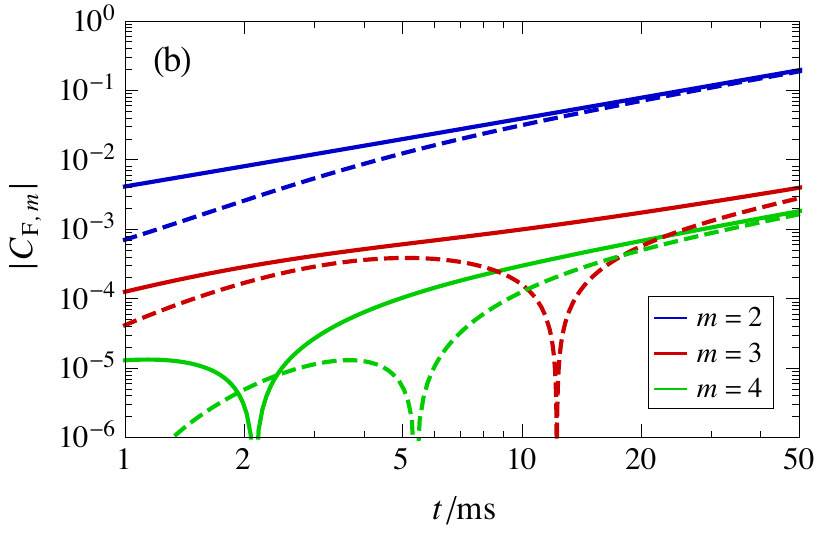}}
\caption{
Factorial cumulants for the single-electron-box system depicted in Fig.~\ref{figure5}. Solid lines show $C_{{\rm F},m}^{\rm in}$ of tunneling-in events and dashed lines represent $C_{{\rm F},m}^{\rm out}$ of tunneling-out events. In (a), the gate charge is chosen $n_{\rm G}=0$ such that the charge states $-1$ and $1$ are degenerate. Detailed balance is fulfilled and thus solid and dashed lines coincide, $C_{{\rm F},m}^{\rm in} = C_{{\rm F},m}^{\rm out}$. In (b), we chose $n_{\rm G}=0.12$. Violation of detailed balance is indicated by $C_{{\rm F},m}^{\rm in} \neq C_{{\rm F},m}^{\rm out}$.}\label{figure6}
\end{figure*}

In Fig.~\ref{figure3}a, we show the current as a function of level energy $\varepsilon$ and bias voltage $V$.
In the white region, the dot remains in its ground state due to Coulomb blockade. In the colored regions, transitions to other states become possible. In most areas of the parameter space, detailed balance holds. However, as depicted in Fig.~\ref{figure3}b, there are also areas where detailed balance is violated, which is indicated by a nonvanishing difference $C_{{\rm F},2}^{\rm in}-C_{{\rm F},2}^{\rm out}$.

In Fig.~\ref{figure4}, we depict the factorial cumulants $C_{{\rm F},m}^{\rm in}(t)$ (solid lines) and $C_{{\rm F},m}^{\rm out}(t)$ (dashed lines) as a function of the orbital level energy $\varepsilon$ for a given bias voltage indicated by the black line in Fig.~\ref{figure3}. For $0< \varepsilon/ U $, double occupation of the quantum dot is suppressed such that the device is a realization of the stochastic model shown in Fig.~\ref{figure1}a, for which detailed balance is expected to be obeyed. Thus, the factorial cumulants are the same for the tunneling-in and tunneling-out statistics.

This is qualitatively different for $-0.3 < \varepsilon/U < 0$. All transitions depicted in Fig.~\ref{figure1}b are possible, except the transition from $\downarrow$ to $\rm d$, i.e., $\Gamma_{{\rm d} \downarrow}=0$. Since both the probabilities $p_\downarrow$ and $p_{\rm d}$ are finite, detailed balance Eq.~$(\ref{detailedB})$ is violated. We find a clear difference between the factorial cumulants of the tunneling-in and the tunneling-out statistics for all cumulants of order $m\ge 2$, see Fig.~\ref{figure4}. Of course, the first cumulants of tunneling in and tunneling out are identical, which is a consequence of the system being in a steady state (note that the first cumulant is equal to the mean number $\langle N \rangle$).

For $-0.7 < \varepsilon/U < -0.3$, also the transition $\downarrow$ to $\rm d$ becomes possible and detailed balance is reestablish. Factorial cumulants are the same for the tunneling-in and tunneling-out statistics.
For even smaller level positions $\varepsilon$, the upper discussion can be repeated due to particle hole symmetry.

We remark that in order to violate detailed balance, it was crucial to break the symmetry between spin $\uparrow$ and $\downarrow$ as well as the symmetry between $0$ and d. Without a Zeeman field~$\Delta=0$ or Coulomb interaction~$U=0$, detailed balance is recovered. Only with a finite Zeeman field and Coulomb interaction, there are the regions in parameter space, as depicted in Fig.~\ref{figure3}b, where detailed balance is violated.

Furthermore, we realize that for the proposed quantum-dot system, a finite bias voltage was applied to achieve a violation of detailed balance, because, otherwise, we could not overcome the charging energy to occupy three different charge states with a finite probability.
There is, however, no fundamental reason to require a bias voltage for breaking detailed balance.
In fact, in the next section, we are going to show an example that already works without bias voltage.

\section{Single-electron box with superconducting lead}\label{sec:box}
Next, we consider a device, see Fig.~\ref{figure5} that has been experimentally realized in Refs.~\cite{saira_environmentally_2010,maisi_real_2011,maisi_full_2014}.
A normal metal single-electron box is weakly tunnel coupled to a superconductor and electrostatically coupled to a normal-conducting single-electron transistor that is sensitive to the charge state of the box. 
At low temperature, only three charge states $-1,0,1$, relative to some reference charge, play a role.
Quasiparticle tunneling induces transitions that change the island's charge by $\pm 1$. 
In addition, there is Andreev tunneling, which directly couples the states $-1$ and $1$ and, therefore, changes the island's charge by $\pm 2$ [as a consequence, counting an Andreev tunneling-in and -out event contributes with a factor of $2$ to $P_N^{\rm in}(t)$ and $P_N^{\rm out}(t)$, respectively].
This device is a realization of the stochastic system sketched in Fig.~\ref{figure1}c, with
\begin{equation}
	{\bf W}^{\rm in}_{z} =
	\left( \begin{array}{ccc}
		-\Gamma_{0 \text{-}1}-\Gamma_{1 \text{-}1} & \Gamma_{\text{-}1 0} & \Gamma_{\text{-}1 1}
\\
		z \Gamma_{0 \text{-}1} & -\Gamma_{\text{-}1 0} -\Gamma_{1 0} & \Gamma_{0 1}
\\
		z^2 \Gamma_{1 \text{-}1} & z\Gamma_{1 0} & -\Gamma_{\text{-}1 1}-\Gamma_{0 1}
	\end{array} \right)
\, .
\end{equation}
Again, the matrix ${\bf W}^{\rm out}_{z}$ is obtained by removing the factors $z$ and $z^2$ from the lower left and adding them to the upper right matrix elements.

The tunneling rates depend on the (normal-state) tunnel resistance, the energies of the different charge states and the coupling to the electromagnetic environment~\cite{saira_environmentally_2010,pekola_environment_2010}.
In order to make close contact to possible experimental realizations, we do not calculate the tunneling rates for some assumed electromagnetic environment, but rather use experimentally measured values taken from Fig.~3b of Ref.~\cite{maisi_real_2011}.
We do this for two different choices of the gate voltage.

First, we consider the symmetric case, $n_{\rm G}=0$, for which the charge states $-1$ and $1$ are energetically degenerate. The transition rates are $\Gamma_\text{0-1}=\Gamma_\text{01}=960 \, {\rm Hz}$, $\Gamma_\text{-10}=\Gamma_\text{10}=9.4 \, {\rm Hz}$, and $\Gamma_\text{1-1}=\Gamma_{\text{-}11}=150 \, {\rm Hz}$. The symmetry guarantees detailed balance, as demonstrated in Fig.~\ref{figure6}a.
This contrasts with the asymmetric case, for which we choose $n_{\rm G}=0.12$. The transition rates are $\Gamma_\text{0-1}= 1280 \, {\rm Hz}$, $\Gamma_\text{01}=650 \, {\rm Hz}$, $\Gamma_\text{-10}=6.5\, {\rm Hz}$, $\Gamma_\text{10}=11.5 \, {\rm Hz}$, $\Gamma_\text{1-1}=630\, {\rm Hz}$, and $\Gamma_\text{-11}=10.8 \, {\rm Hz}$.
In this case, the factorial cumulants of order $m\ge 2$ for tunneling in and tunneling out differ from each other, see Fig.~\ref{figure6}b. 
The dips shown in Fig.~\ref{figure6}b indicate sign changes of $C_{{\rm F},m}^{\rm out}$ and $C_{{\rm F},m}^{\rm in}$.
They occur independently of each other.
At long times, e.g. $t=50\, \rm ms$, the signs of $C_{{\rm F},m}^{\rm out}$ and $C_{{\rm F},m}^{\rm in}$ coincide.

We emphasize that, for this device, Andreev tunneling is crucial to establish steady states that violate detailed balance.
Without this charge-transfer channel, there would be no possibility for a finite net probability current around a closed loop.

\section{Conclusion}
Time-resolved measurements of electron tunneling into and out of a quantum dot or a single-electron box in the steady-state limit define a suitable tool to test whether or not detailed balance is fulfilled.
To generate a steady state that violates detailed balance, one needs to realize a stochastic model with at least three different states and tunneling rates that allow for a finite net probability current around a closed loop.
We have suggested two devices in which such detailed-balance-violating steady states can be established: a spin-split, single-level quantum dot at large bias voltage and a metallic single-electron box coupled to a superconductor.
Both suggestions seem experimentally feasible with nowadays technology.

\begin{acknowledgement}
We acknowledge financial support from DFG via KO 1987/5 and SFB 1242.
\end{acknowledgement}

\end{document}